\documentclass[conference]{IEEEtran}
\IEEEoverridecommandlockouts
\usepackage{cite}
\usepackage{amsmath,amssymb,amsfonts}
\usepackage{algorithmic}
\usepackage{graphicx}
\usepackage{textcomp}
\usepackage{xcolor}
\usepackage{hyperref}
\usepackage{tabularx}
\usepackage{multirow}
\usepackage{subcaption}
\usepackage{pifont}
\newcommand{\cmark}{\ding{51}}%
\newcommand{\xmark}{\ding{55}}%
\def\BibTeX{{\rm B\kern-.05em{\sc i\kern-.025em b}\kern-.08em
    T\kern-.1667em\lower.7ex\hbox{E}\kern-.125emX}}
\begin{document}

\title{Leveraging LLMs for the Quality Assurance of Software Requirements
\thanks{The presented work has been developed within the research project \textsc{Streamdiver}, which is funded by the Austrian Research Promotion Agency (FFG) under the project number \textsc{886205}.}
}

\author{\IEEEauthorblockN{1\textsuperscript{st} Sebastian Lubos}
\IEEEauthorblockA{\textit{Graz University of Technology, Austria}\\
slubos@ist.tugraz.at}
\and
\IEEEauthorblockN{2\textsuperscript{nd} Alexander Felfernig}
\IEEEauthorblockA{\textit{Graz University of Technology, Austria}\\
afelfern@ist.tugraz.at}
\and
\IEEEauthorblockN{3\textsuperscript{rd} Thi Ngoc Trang Tran}
\IEEEauthorblockA{\textit{Graz University of Technology, Austria}\\
ttrang@ist.tugraz.at}
\and
\IEEEauthorblockN{4\textsuperscript{th} Damian Garber}
\IEEEauthorblockA{\textit{Graz University of Technology, Austria}\\
dgarber@ist.tugraz.at}
\and
\IEEEauthorblockN{5\textsuperscript{th} Merfat El Mansi}
\IEEEauthorblockA{\textit{Graz University of Technology, Austria}\\
merfat.el-mansi@student.tugraz.at}
\and
\IEEEauthorblockN{6\textsuperscript{th} Seda Polat Erdeniz}
\IEEEauthorblockA{\textit{Graz University of Technology, Austria}\\
sedapolat@gmail.com}
\and
\IEEEauthorblockN{\phantom{-----------------------------------------------}7\textsuperscript{th} Viet-Man Le}
\IEEEauthorblockA{\phantom{-------------------------------------------------------}\textit{Graz University of Technology, Austria}\\
\phantom{------------------------------------------------------}vietman.le@ist.tugraz.at}
}

\maketitle

\begin{abstract}
Successful software projects depend on the quality of software requirements. Creating high-quality requirements is a crucial step toward successful software development. Effective support in this area can significantly reduce development costs and enhance the software quality. In this paper, we introduce and assess the capabilities of a \textit{Large Language Model (LLM)} to evaluate the quality characteristics of software requirements according to the \textit{ISO 29148} standard. We aim to further improve the support of stakeholders engaged in \textit{requirements engineering (RE)}. We show how an LLM can assess requirements, explain its decision-making process, and examine its capacity to propose improved versions of requirements. We conduct a study with software engineers to validate our approach. Our findings emphasize the potential of LLMs for improving the quality of software requirements.
\end{abstract}


\section{Introduction}
The success of software projects depends on the quality of software requirements \cite{kamata2007how}. Formulating high-quality requirements constitutes an essential step towards achieving favorable outcomes \cite{azham2016role}. However, reaching such quality requires significant collaborative efforts from stakeholders engaged in the project. To assist this process, various tools and techniques have demonstrated their efficiency in supporting multiple related tasks \cite{felfernig2013overview}, including the classification \cite{kurtanovic2017automatically}, prioritization \cite{felfernig2018towards,vijayakumar2021use}, and quality assessment of requirements \cite{parra2015methodology}.

A recent related development in this direction has emerged with the introduction of powerful \textit{Large Language Models (LLMs)} \cite{min2023llm} such as \textit{GPT-4} \cite{openai2023gpt4} and \textit{Llama 2} \cite{touvron2023llama}. These models show remarkable capabilities in generating natural language, producing a huge variety of accurately crafted responses to input prompts. Applying these models in requirements engineering has received increasing attention in past years. Among others, they have been explored for tasks such as inconsistency detection within sets of requirements \cite{fantechi2023inconsistency,bertram2023leveraging} and identification of incomplete requirements \cite{luitel2023using}.

While the existing literature explores interesting possibilities for leveraging LLMs to assess and improve specific quality aspects of requirements, addressing the broader issue of \emph{evaluating} the overall quality of software requirements, including multiple quality characteristics, remains unexplored. The study presented in this paper investigates the usage of an LLM to assess whether individual requirement statements adhere to the set of quality characteristics defined in \textit{ISO 29148} \cite{iso29148}. We outline our approach to instruct the LLM in this assessment task and investigate its capacity to transparently explain decisions, aiming to enhance trust in the system. Furthermore, we examine the LLM's potential to propose improved versions of requirements to correct identified quality flaws. To assess the effectiveness of this approach, we present the findings of an empirical user study conducted with participants having a background in software engineering, offering a proof-of-concept evaluation of the approach's benefits.

With this work, we provide a novel approach to using an LLM to transparently evaluate the quality of software requirements by explaining the decisions and presenting suggestions for improvement. This includes three primary contributions. \emph{Firstly}, we show how an LLM can be instructed to evaluate the quality of software requirement statements, comparing its accuracy in flaw detection with that of study participants. \emph{Secondly}, we analyze the LLM's capacity to consistently explain its assessments, thereby enhancing transparency and fostering trust in the evaluation. \emph{Thirdly}, we investigate the LLM's capability to offer constructive suggestions for improved requirements addressing identified flaws.

The remainder of the paper is organized as follows. Section \ref{sec:background} covers essential aspects required to follow the paper. This includes requirement quality characteristics as defined in \textit{ISO 29148} and fundamental knowledge related to applying LLMs in evaluating software requirements. Section \ref{sec:related_work} gives an overview of existing work on the application of LLM-enhanced techniques in the context of software requirements engineering. In Section \ref{sec:evaluation}, we present the research questions, outline the experimental setup of the user study, and provide a comprehensive discussion of its results. Section \ref{sec:discussion} explores the implications of the results, taking into account the limitations of this study and suggesting open issues for future work. Finally, Section \ref{sec:conclusion} summarizes and concludes the paper.

\section{Background} \label{sec:background}
\subsection{Quality Characteristics of Software Requirements}\label{ssec:quality_characteristics}
\textit{Requirements engineering (RE)} is an important phase in the software development process, and the creation of high-quality requirements is crucial for successful software development \cite{azham2016role}. The \textit{ISO 29148} \cite{iso29148} describes the international standard for systems and software engineering, including RE. This standard defines \textit{nine} characteristics (dimensions) for individual software requirements, describing the capabilities, characteristics, constraints, and quality factors of a software system. Table \ref{tab:characteristics} summarizes these quality characteristics.

\begin{table}[h!]
    \centering
    \begin{tabular}{c|m{0.34\textwidth}}
         \textbf{Characteristic} & \textbf{Description} \\ \hline\hline
         Appropriate    &  The level of abstraction is adequate, excludes unnecessary constraints, and avoids implementation details. \\ \hline
         Complete   &  All information needed to understand the requirement is included in the description. \\ \hline
         Conforming &  The representation of the requirement follows an approved standard template.  \\ \hline
         Correct    &  The need is accurately represented in the requirement. \\ \hline
         Feasible   &  The requirement is realizable within the given system constraints considering an acceptable risk. \\ \hline
         Necessary  &  The requirement defines an essential aspect of the system and is irremovable without causing a deficiency. \\ \hline
         Singular   &  The requirement defines only one aspect of the system. \\ \hline
         Unambiguous    &   The requirement is clearly stated, understandable, and allows only one interpretation.  \\ \hline
         Verifiable &  The requirement is formulated in a way that its fulfillment can be proven or, in the best case, measured. \\
    \end{tabular}
    \caption{Characteristics of high-quality software requirements as defined in \textit{ISO 29148} \cite{iso29148}.}
    \label{tab:characteristics}
    \vspace{-8pt}
\end{table}

To determine if project requirements fulfill these quality characteristics, a review by project stakeholders is required. These stakeholders need to understand the project's scope and have experience in software projects. Their task is to decide whether the software requirements meet the mentioned quality characteristics. This evaluation process can involve multiple stakeholders reviewing the requirements and reaching a consensus through a majority decision.

\subsection{Instructing LLMs to Evaluate Software Requirements}\label{ssec:instructing_llms}
A \textit{Large Language Model (LLM)} is an advanced \textit{Natural Language Processing (NLP)} model, demonstrating capabilities in comprehending and generating human-like text \cite{min2023llm}. These models find application across diverse domains, such as content generation \cite{moore2023empowering}, language translation \cite{huang2023towards}, and code completion \cite{hou2023large}, enabling the creation of relevant information based on contextual input. To execute various tasks, LLMs are provided with specific input queries known as \textit{prompts}, guiding them to generate desired outputs or responses in natural language \cite{reynolds2021prompt,beurer-kellner2023prompting}.

\begin{figure}[h!]
    \centering
    \includegraphics[width=\linewidth]{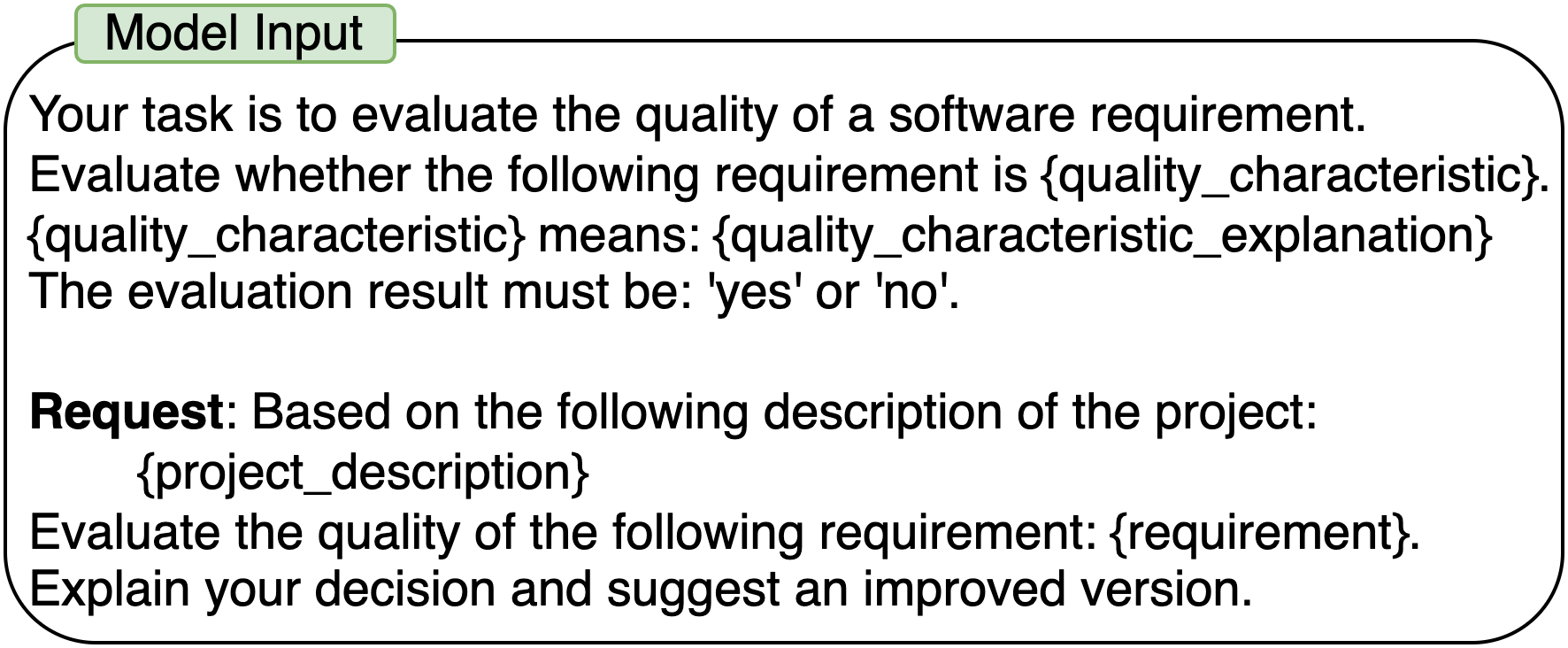}
    \caption{LLM prompt template used to evaluate a software requirement for a specified quality characteristic and project. Variables are written in curly brackets ``\{\dots\}''.}
    \label{fig:req-evaluation}
    \vspace{-16pt}
\end{figure}

When assessing individual software requirements, the LLM must understand the project scope. To achieve this, the project scope description must be incorporated into the prompt. The LLM then has to be instructed to evaluate a given requirement, considering both, the project description and the definition of a specified quality characteristic (refer to Table \ref{tab:characteristics}), as additional contextual information. The primary task of the LLM is then to classify whether the requirement satisfies the specified quality dimension and to explain its decision. If a quality issue is identified, the LLM is further instructed to propose an improved version of the requirement that addresses the identified flaw. The related LLM prompt used in our study is depicted in Figure \ref{fig:req-evaluation}, including the quality characteristic to be evaluated, its definition, and the project scope description. Additionally, the model's output space is limited to two options, indicating whether the requirement meets the quality criteria or not.

\section{Related Work} \label{sec:related_work}
The integration of AI-enhanced tools for requirements engineering has received increased attention in recent years. The application of \textit{machine learning (ML)} and NLP forms a crucial focus of research in this domain. Cheeliger et al. \cite{cheligeer2022machine} conducted a comprehensive literature review on machine learning in requirements engineering, categorizing publications into four tasks: preparation, collection, validation, and negotiation. The review's findings suggest that while the potential of ML in this field is promising, its seamless integration into existing engineering workflows remains challenging. Our work seeks to contribute additional value, specifically in the validation task, by facilitating the quality assessment process.

Leveraging LLMs in software requirement engineering is not novel. Arora et al. \cite{arora2023advancing} conducted a preliminary evaluation of using an LLM to elicitate software requirements for a real-world application. The study results emphasized the feasibility of LLMs for this task. Fantechi et al. \cite{fantechi2023inconsistency} explored the use of \textit{ChatGPT} to identify conflicting requirements. The results of the study suggest that the model cannot replace human judgment but can complement manual analysis and speed up the process. In a similar direction, Bertram et al. \cite{bertram2023leveraging} incorporated an LLM into a suggested toolchain to detect inconsistencies in large requirement specifications within the automotive industry. The LLM translated the natural language requirements into structured English, enabling formal processing for consistency checking. Luitel et al. \cite{luitel2023using} investigated the potential of LLMs in enhancing the completeness of requirements expressed in natural language. The evaluation indicates that LLMs can assist in identifying incompleteness within requirements.

Our objective is to enhance the understanding of LLMs in identifying quality issues within software requirements with regard to various dimensions. In exploring the potential of LLMs, we investigate the applicability of the model to provide explainable insights into its assessments, enhance the transparency of decisions, and increase trust in the model. Additionally, we investigate the model's capability to offer constructive suggestions for improving flawed requirements.

\section{Quality Assurance of Software Requirements with LLMs} \label{sec:evaluation}
\subsection{Methodology} \label{sec:methodology}
Considering the rich capabilities of LLMs in comprehending contexts \cite{zhou2023context} and reasoning over textual descriptions \cite{huang2023reasoning}, we expect their potential to support requirements engineering tasks. We expect these models to support the identification of potential weaknesses in requirements artifacts and, ideally, locate quality issues precisely, offering explanations and suggestions for improvement. In doing so, these models might serve as valuable assistants, potentially reducing the need for extensive review cycles.

To evaluate our assumptions, we formulate the following hypotheses:
\begin{itemize}
    \item [\textbf{H1}:] LLMs achieve comparable performance to software engineers in identifying quality issues in software requirements concerning various quality characteristics.
    \item [\textbf{H2}:] LLMs can reasonably explain identified quality issues.
    \item [\textbf{H3}:] LLMs can suggest improved requirement statements, enhancing their quality.
\end{itemize}

Building upon these hypotheses, we define the following research questions:
\begin{itemize}
    \item [\textbf{R1}:] How accurately can LLMs identify quality issues in software requirements across different quality characteristics?
    \item [\textbf{R2}:] Can LLMs provide meaningful explanations for their quality assessments of software requirements?
    \item [\textbf{R3}:] Do LLMs have the ability to suggest improved versions of software requirements, addressing identified flaws?
\end{itemize}

\subsection{Study Participants}
Our empirical study involved individuals with educational and professional backgrounds in software engineering. While participants were not required to be experienced experts in RE, they needed practical experience working on software projects. This background of participants warranted that participants had fundamental knowledge in working with software requirements. The study instructions included a definition and explanation of the quality characteristics to ensure a clear understanding of those among all participants.

\subsection{Dataset and Preprocessing} \label{ssec:dataset}
For our empirical study, we used two example projects described in the following:

\subsubsection{Stopwatch Project}
\begin{figure}
    \centering
    \includegraphics[width=\linewidth]{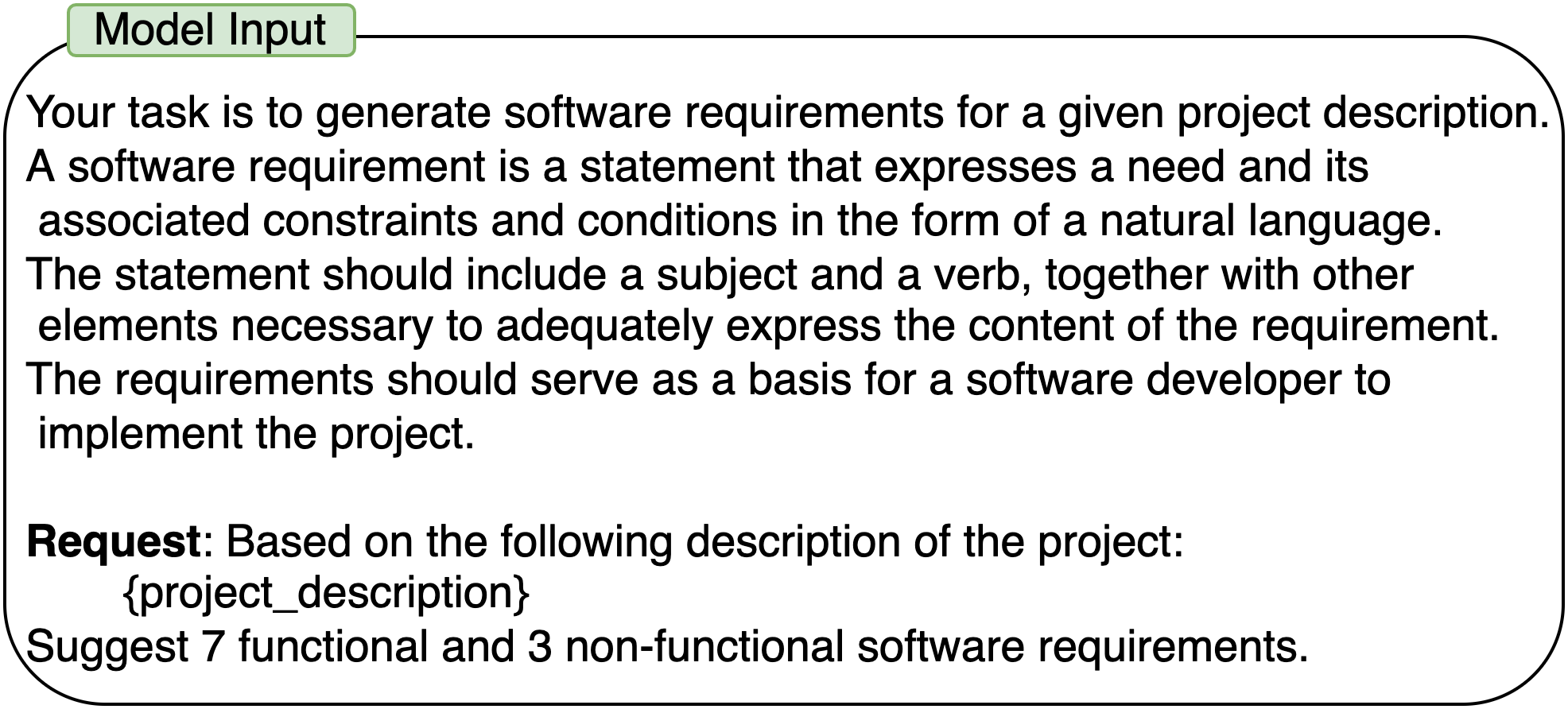}
    \caption{LLM prompt template used to generate software requirements for an example project to implement a \textit{Stopwatch app for Android smartphones}. Variables are written in curly brackets ``\{\dots\}''.}
    \label{fig:req-generation}
    \vspace{-16pt}
\end{figure}

We initially considered a hypothetical small project. As most software projects are typically large and complex, understanding their context and evaluating their requirements requires high effort. For this reason, we decided to first consider a relatively simple project comprising ten requirements, seven functional and three non-functional, to evaluate our proof-of-concept. The software to be described with those requirements is a \emph{Stopwatch app designed for Android smartphones}. We used an LLM\footnote{\textit{Llama 2} \cite{touvron2023llama} with 70 billion parameters as described in Section \ref{sec:llm-req-evaluation}.} to generate this set of requirements by using the prompt shown in Figure \ref{fig:req-generation}, along with the following project scope description: ``\textit{Develop an intuitive and user-friendly Stopwatch app for Android smartphones that allows users to easily measure time intervals with precision. The app should have a simple and clean interface, allowing users to start, pause, and reset the stopwatch easily. Additionally, the app should display the total time elapsed and provide options for split and lap times. The goal is to create an efficient and reliable tool that can be used in various contexts such as sports, cooking, or any other activity where accurate time measurement is important}.''

The LLM was not explicitly instructed to fulfill specific quality characteristics during the generation of requirements, which means that realistic and potentially flawed samples were generated. To ensure their coherence and suitability as an explanatory example for our study, we carefully reviewed the generated requirements\footnote{We deliberately refrained from correcting the requirements concerning the quality characteristics outlined in Table \ref{tab:characteristics}, as we wanted to consider a realistic example that could contain quality issues.}. The generated requirements are shown in Table \ref{tab:requirements-stopwatch}

\begin{table}
    \centering
    \begin{tabular}{c|m{0.4\textwidth}}
        \textbf{Id} & \textbf{Requirement} \\ \hline \hline
        $\mathbf{r_1}$ & The app shall allow users to start the stopwatch by tapping a prominent 'Start' button. \\ \hline
        $\mathbf{r_2}$ & The app shall allow users to pause the stopwatch by tapping a prominent 'Pause' button. \\ \hline
        $\mathbf{r_3}$ & The app shall allow users to reset the stopwatch to zero by tapping a prominent 'Reset' button. \\ \hline
        $\mathbf{r_4}$ & The app shall display the total time elapsed since the last reset. \\ \hline
        $\mathbf{r_5}$ & The app shall provide an option for split times, allowing users to manually enter a split time and display it alongside the total time elapsed. \\ \hline
        $\mathbf{r_6}$ & The app shall provide an option for lap times, allowing users to manually enter a lap time and display it alongside the total time elapsed. \\ \hline
        $\mathbf{r_7}$ & The app shall allow users to view their previous splits and laps, including the time taken for each split and lap. \\ \hline
        $\mathbf{r_8}$ & The app shall be designed with a simple and clean interface, ensuring ease of use for users of all ages and skill levels. \\ \hline
        $\mathbf{r_9}$ & The app shall be optimized for performance, ensuring that it operates efficiently and without significant lag or errors. \\ \hline
        $\mathbf{r_{10}}$& The app shall be compatible with Android smartphones running version 10 or higher, ensuring that it can be installed and used on a wide range of devices. \\
    \end{tabular}
    \caption{Generated requirements for the development of a \textit{Stopwatch app for Android smartphones}.}
    \label{tab:requirements-stopwatch}
    \vspace{-16pt}
\end{table}

\subsubsection{DigitalHome Project}
As a second example, we focused on the software requirements of a real-world project included in the \textit{PURE dataset} \cite{ferrari2018pure}. This dataset collects requirement documents from public projects. From those samples, we selected \textit{DigitalHome}, a project detailing the requirements regarding a smart home prototype, as a representative case for our study. These requirements need to be evaluated by a group of study participants who understand the project scope, and we assumed that a basic understanding of smart homes can be presumed. The project includes \textit{63} requirements, comprising \textit{42 functional} and \textit{21 non-functional} requirements.

\subsection{LLM-based Requirement Evaluation} \label{sec:llm-req-evaluation}
We conducted an LLM-based evaluation of requirements utilizing the \textit{Llama 2} language model \cite{touvron2023llama} with 70 billion parameters, fine-tuned to complete chat responses. We decided to perform our experiments with this LLM, as it achieved competitive benchmark performance\footnote{\url{https://huggingface.co/spaces/HuggingFaceH4/open_llm_leaderboard}} at the time the study was conducted and was published as open-source\footnote{\url{https://github.com/meta-llama/llama}}. For practicality, we accessed the hosted model on \textit{Replicate}\footnote{Model version: \textit{meta/llama-2-70b-chat:\\02e509c789964a7ea8736978a43525956ef40397be9033abf9fd2badfe68c9e3}} with parameters $\mathrm{temperature=0.01}, \mathrm{max\_new\_tokens=2000}$. Given that real-time responses were not a criterion for our evaluation, we used the most accurate model despite its slower response generation speed (\textit{$\approx$15-30 seconds}). Details on the LLM instructions for this task are provided in Section \ref{ssec:instructing_llms}.

\subsection{Experimental Setup} \label{sec:experimental_setup}
To assess the validity of our hypotheses regarding the capabilities of LLMs in evaluating the quality of software requirements and to address the research questions outlined in Section \ref{sec:methodology}, we implemented the following procedure for our experiments:
\begin{enumerate}
    \item Instruct study participants to evaluate the requirements of both example projects (see Section \ref{ssec:dataset}) based on the \textit{ISO 29148} quality characteristics (see Section \ref{ssec:quality_characteristics}) and their project scope description.
    \item Use the requirements of both example projects as input for LLM requests, instructing it to assess their quality (see Section \ref{sec:llm-req-evaluation}) and record the model's responses. The project scope descriptions and quality characteristic definitions are part of the LLM prompt.
    \item Instruct study participants to evaluate the LLM responses, considering:
        \begin{itemize}
            \item The \emph{agreement} with the LLM's decision (\textbf{R1})
            \item The \emph{plausibility} of the LLM's explanation (\textbf{R2})
            \item The \emph{quality} of the improved requirement suggested by the LLM (\textbf{R3})
        \end{itemize}
    \item Compare the assessment of quality flaws by the LLM and study participants, using the study participants' responses as the baseline.
\end{enumerate}

Each requirement and its corresponding LLM response has been evaluated by at least \textit{four} individuals with software engineering backgrounds. They used the project description and an explanation of quality characteristics as a guideline for their assessments. We used their evaluation to establish the ground truth for our project examples. A majority vote was applied to label the samples, determining if a requirement meets the observed quality characteristic. If a majority consensus was not reached or if the majority expressed uncertainty, we labeled the requirement as not fulfilling the observed quality characteristic. Our assumption was that when a group of reviewers cannot reach a unanimous decision, the requirement is flawed, and additional review is needed.

Using the described process, we had two phases of study participants evaluating the quality characteristics of requirements. In the first evaluation phase, which we name \textit{independent assessment}, the study participant classified requirements without knowledge about the decision of the LLM. In the second phase, the study participants knew the decision and explanation of the LLM and decided whether they agreed. We refer to the second phase as \textit{bound assessment}. Study participants might have classified requirements differently in the second phase, indicating that the LLM has convinced them. During this phase, study participants evaluated the explanation provided by the LLM by categorizing it as either plausible, implausible, or neutral. Similarly, the quality of suggested requirement improvements by the LLM was evaluated.

\subsection{Results}
\subsubsection{Agreement between Study Participants and LLM (R1)}
As a first step, we evaluate the agreement between the LLM and study participants when assessing the quality characteristics of requirements in our investigated \textit{Stopwatch} and \textit{DigitalHome} projects. In line with our initial hypothesis, we anticipate a high level of agreement between the two. To measure this agreement, we employ the \textit{Cohen's Kappa} metric \cite{mchugh2012interrator}, a measure of inter-rater reliability that quantifies the extent to which two raters agree on the categorical labeling of data samples, accounting for the potential of agreement by chance. We considered an agreement between the study participants and LLM when their assessment of a requirement and quality characteristic was identical. 

In Table \ref{tab:cohen}, the \textit{Cohen's Kappa} values are shown. For the independent assessment, a \emph{weak} agreement is observed between the LLM and study participants' evaluations for the \textit{Stopwatch} project ($0.4 <= \kappa <0.6 $), according to the interpretation outlined by McHugh in \cite{mchugh2012interrator}. In the bound assessment, we could measure a \emph{substantial} agreement ($0.6 <= \kappa <0.8 $) between the LLM and study participants for the \textit{Stopwatch} example, which indicates that the study participants agreed with the LLM assessment of requirements in many cases. For the \textit{DigitalHome} project, we could not establish an agreement in the independent assessment, while a \emph{fair} agreement ($0.2 <= \kappa <0.4 $) was identified in the bound assessment scenario. The metric values show that we cannot assert that the LLM and study participants yield identical evaluations.

\begin{table}
    \centering
    \begin{tabular}{c|c|c}
         \textbf{Project} & \textbf{Independent Assessment} & \textbf{Bound Assessment} \\ \hline \hline
         \textit{Stopwatch} & 0.4028 & 0.7545 \\
         \textit{DigitalHome} & 0.0486 & 0.2223
    \end{tabular}
    \caption{Cohen's Kappa value indicating the inter-rater agreement between the study participants and LLM in classifying requirements as flawed. Independent assessment means that both raters answered without knowledge about how the other rater answered, while bound assessment describes the situation where the study participants classified the requirements being aware of the decision and explanation of the LLM.}
    \label{tab:cohen}
    \vspace{-16pt}
\end{table}

While \textit{Cohen's Kappa} metric measures agreement between raters, it lacks an assessment of the validity of responses, i.e., whether two raters have correctly classified samples. To understand the values better, we provide an additional perspective to the evaluations. The Tables \ref{tab:participants-decision-dimension-requirements-stopwatch} and \ref{tab:llm-decision-dimension-requirements-stopwatch} show the assessments of quality characteristics by study participants and the LLM for the requirements of the \textit{Stopwatch} project (refer to Table \ref{tab:requirements-stopwatch}). The first table indicates the majority decisions (see Section \ref{sec:experimental_setup}) of study participants if a requirement fulfills a quality characteristic or not. The second table shows the LLM decisions for the same requirements. The sum values $\mathbf{\Sigma_{req}}$ in the tables reflect the number of fulfilled quality characteristics per requirement. The sum in column $\mathbf{\Sigma_{qc}}$ shows the number of requirements fulfilling a specific quality characteristic.

\begin{table}
    \begin{subtable}{\linewidth}
        \centering
        \caption{Study participants' assessment.}
        \label{tab:participants-decision-dimension-requirements-stopwatch}
        \begin{tabular}{>{\centering\arraybackslash}m{0.16\textwidth}|>{\centering\arraybackslash}m{0.014\textwidth}>{\centering\arraybackslash}m{0.014\textwidth}>{\centering\arraybackslash}m{0.014\textwidth}>{\centering\arraybackslash}m{0.014\textwidth}>{\centering\arraybackslash}m{0.014\textwidth}>{\centering\arraybackslash}m{0.014\textwidth}>{\centering\arraybackslash}m{0.014\textwidth}>{\centering\arraybackslash}m{0.014\textwidth}>{\centering\arraybackslash}m{0.014\textwidth}>{\centering\arraybackslash}m{0.04\textwidth}|>{\centering\arraybackslash}m{0.04\textwidth}}
                                  & $\mathbf{r_1}$ & $\mathbf{r_2}$ & $\mathbf{r_3}$ & $\mathbf{r_4}$& $\mathbf{r_5}$ & $\mathbf{r_6}$ & $\mathbf{r_7}$ & $\mathbf{r_8}$ & $\mathbf{r_9}$ & $\mathbf{r_{10}}$ & $\mathbf{\Sigma_{qc}}$ \\ \hline
             \textit{Appropriate} & \xmark & \xmark & \xmark & \cmark & \cmark & \cmark & \cmark & \cmark & \cmark & \cmark & \textbf{7} \\
             \textit{Complete}    & \xmark & \xmark & \cmark & \xmark & \xmark & \xmark & \xmark & \xmark & \xmark & \cmark & \textbf{2} \\
             \textit{Conforming}  & \cmark & \xmark & \cmark & \xmark & \cmark & \xmark & \cmark & \cmark & \xmark & \cmark & \textbf{6} \\
             \textit{Correct}     & \cmark & \cmark & \cmark & \cmark & \cmark & \xmark & \cmark & \cmark & \cmark & \cmark & \textbf{9} \\
             \textit{Feasible}    & \cmark & \cmark & \cmark & \cmark & \cmark & \cmark & \cmark & \cmark & \cmark & \cmark & \textbf{10} \\
             \textit{Necessary}   & \cmark & \cmark & \cmark & \cmark & \cmark & \cmark & \cmark & \cmark & \cmark & \cmark & \textbf{10} \\
             \textit{Singular}    & \cmark & \cmark & \cmark & \cmark & \cmark & \cmark & \cmark & \xmark & \xmark & \xmark & \textbf{7} \\
             \textit{Unambiguous} & \cmark & \cmark & \cmark & \xmark & \xmark & \xmark & \cmark & \xmark & \xmark & \cmark & \textbf{5} \\
             \textit{Verifiable}  & \cmark & \cmark & \cmark & \cmark & \xmark & \xmark & \xmark & \xmark & \cmark & \cmark & \textbf{6} \\ \hline
             $\mathbf{\Sigma_{req}}$         & \textbf{7} & \textbf{6} & \textbf{8} & \textbf{6} & \textbf{6} & \textbf{4} & \textbf{7} & \textbf{5} & \textbf{5} & \textbf{8} & \textbf{62} \\
        \end{tabular}
    \end{subtable}
    \begin{subtable}{\linewidth}
        \centering
        \caption{LLM assessment.}
        \label{tab:llm-decision-dimension-requirements-stopwatch}
        \begin{tabular}{>{\centering\arraybackslash}m{0.16\textwidth}|>{\centering\arraybackslash}m{0.014\textwidth}>{\centering\arraybackslash}m{0.014\textwidth}>{\centering\arraybackslash}m{0.014\textwidth}>{\centering\arraybackslash}m{0.014\textwidth}>{\centering\arraybackslash}m{0.014\textwidth}>{\centering\arraybackslash}m{0.014\textwidth}>{\centering\arraybackslash}m{0.014\textwidth}>{\centering\arraybackslash}m{0.014\textwidth}>{\centering\arraybackslash}m{0.014\textwidth}>{\centering\arraybackslash}m{0.04\textwidth}|>{\centering\arraybackslash}m{0.04\textwidth}}
                                  & $\mathbf{r_1}$ & $\mathbf{r_2}$ & $\mathbf{r_3}$ & $\mathbf{r_4}$& $\mathbf{r_5}$ & $\mathbf{r_6}$ & $\mathbf{r_7}$ & $\mathbf{r_8}$ & $\mathbf{r_9}$ & $\mathbf{r_{10}}$ & $\mathbf{\Sigma_{qc}}$ \\ \hline
             \textit{Appropriate} & \xmark & \xmark & \xmark & \cmark & \xmark & \xmark & \xmark & \cmark & \xmark & \cmark & \textbf{3} \\
             \textit{Complete}    & \xmark & \xmark & \xmark & \cmark & \xmark & \xmark & \xmark & \xmark & \xmark & \cmark & \textbf{2} \\
             \textit{Conforming}  & \cmark & \xmark & \xmark & \cmark & \xmark & \xmark & \cmark & \cmark & \xmark & \cmark & \textbf{5} \\
             \textit{Correct}     & \xmark & \xmark & \xmark & \cmark & \cmark & \xmark & \cmark & \cmark & \cmark & \cmark & \textbf{6} \\
             \textit{Feasible}    & \cmark & \cmark & \cmark & \cmark & \cmark & \cmark & \cmark & \cmark & \cmark & \cmark & \textbf{10} \\
             \textit{Necessary}   & \cmark & \cmark & \cmark & \cmark & \cmark & \cmark & \cmark & \cmark & \cmark & \cmark & \textbf{10} \\
             \textit{Singular}    & \xmark & \xmark & \xmark & \xmark & \xmark & \xmark & \xmark & \xmark & \xmark & \xmark & \textbf{0} \\
             \textit{Unambiguous} & \xmark & \xmark & \xmark & \cmark & \xmark & \xmark & \xmark & \xmark & \xmark & \cmark & \textbf{2} \\
             \textit{Verifiable}  & \xmark & \xmark & \xmark & \cmark & \xmark & \xmark & \xmark & \xmark & \xmark & \cmark & \textbf{2} \\ \hline
             \textit{$\mathbf{\Sigma_{req}}$}         & \textbf{3} & \textbf{2} & \textbf{2} & \textbf{8} & \textbf{3} & \textbf{2} & \textbf{4} & \textbf{5} & \textbf{4} & \textbf{8} & \textbf{40} \\
        \end{tabular}
    \end{subtable}
    \caption{Requirements evaluation of the \textit{Stopwatch} project. ``\cmark'' indicates a requirement that fulfills the quality characteristic. $\mathbf{\Sigma_{req}}$ shows the number of fulfilled quality characteristics per requirement. $\mathbf{\Sigma_{qc}}$ shows the number of requirements fulfilling a quality characteristic.}
    \label{tab:decision-dimension-requirements-stopwatch}
    \vspace{-6pt}
\end{table}

The evaluation summary reveals relevant findings. \emph{Firstly}, the total count of requirements fulfilling a quality characteristic is higher for study participants, suggesting that the LLM tends to evaluate more requirements negatively. The observation can be confirmed for the \textit{DigitalHome} project (see Table \ref{tab:decision-digitalhome}), where the study participants decided on more than twice as many assessments (\textit{514}) that a requirement fulfilled the characteristic, compared to the LLM (\textit{225}).
\emph{Secondly}, substantial differences exist in the positive evaluations of various quality dimensions, with the LLM notably assessing fewer requirements as \emph{appropriate}, \emph{correct}, \emph{singular}, \emph{unambiguous}, and \emph{verifiable}. Again, the same observation was made for the \textit{DigitalHome} project. Table \ref{tab:decision-digitalhome} shows the sum of requirements fulfilling a quality characteristic given the study participants' and LLM's decisions.

\begin{table}[t!]
    \centering
    \begin{tabular}{c|c|c}
         & \textbf{Study Participants} & \textbf{LLM} \\ \hline \hline
        \textit{appropriate} & 60 &  12 \\
        \textit{complete}    & 53 &  15 \\
        \textit{conforming}  & 61 &  42 \\
        \textit{correct}     & 63 &  27 \\
        \textit{feasible}    & 61 &  56 \\
        \textit{necessary}   & 59 &  48\\
        \textit{singular}    & 57 &  5 \\
        \textit{unambiguous} & 49 &  18 \\
        \textit{verifiable}  & 51 &  2 \\ \hline
        $\mathbf{\Sigma_{reqs}}$  & \textbf{514} &  \textbf{225} \\
    \end{tabular}
    \caption{The number of requirements in the \textit{DigitalHome} project fulfilling a quality characteristic based on the assessment of the study participants and LLM evaluation. $\mathbf{\Sigma_{reqs}}$ is the sum of fulfilled characteristics of requirements. The total number of evaluated requirements in this project is \textit{63}.}
    \label{tab:decision-digitalhome}
    \vspace{-16pt}
\end{table}


To assess the validity of the LLM evaluation, we establish the study participants' assessment as the ground truth. This information is used to calculate the \emph{precision} and \emph{recall} of the LLM's capacity to identify quality flaws in requirements accurately, using the equations \ref{eq:precision} and \ref{eq:recall}.

\begin{equation} \label{eq:precision}
    Precision = \frac{\#Flawed~reqs.~identified~by~LLM}{\#Reqs.~classified~as~flawed~by~LLM}
\end{equation}

\begin{equation} \label{eq:recall}
    Recall = \frac{\#Flawed~reqs.~identified~by~LLM}{\#All~flawed~reqs.}
\end{equation}

In this context, the \emph{precision} describes the proportion of identified requirements with quality issues among all requirements the LLM has classified as flawed. The \emph{recall} values represent how many requirements with quality issues in the ground truth have been identified correctly by the LLM. The values for the independent and bound assessments are reported in Table \ref{tab:req_evaluation}.

For the independent assessment, the precision is relatively low in both investigated projects, which implies a higher likelihood of false positives. Quality issues seem to be incorrectly recognized by the LLM using the study participants' evaluation as ground truth. However, the precision value is higher when reviewing the bound assessment. This can be explained by the fact that the study participants' decisions have changed regarding the LLM assessment and explanation. We assume that the study participants received a new perspective on the requirement and, therefore, reconsidered their decision. This means manual verification is still required, even if the number of misclassified requirements is lower. For example, in the \textit{Stopwatch} project, the LLM incorrectly identified \textit{25} out of \textit{90} items as flawed during the independent assessment. This number decreased to \textit{5} in the bound assessment. Considering the precision values for the \textit{DigitalHome} project, a trend to higher precision in the bound assessment can be observed. Yet, the values are much lower, as the number of wrongly identified flaws by the LLM is quite high. We assume that this is related to the project scope, which is more complex than the \textit{Stopwatch} example.

For both projects and assessment phases, high recall values were measured. Those affirm that a large number of quality flaws are indeed identified by the LLM. This outcome suggests that the LLM can be effectively utilized to highlight quality issues in software requirements and guide stakeholders toward requirements that may need improvement.

\begin{table}[t!]
    \centering
    \begin{tabular}{c|cc|cc}
         & \multicolumn{2}{c|}{\textit{Independent Assessment}} & \multicolumn{2}{c}{\textit{Bound Assessment}} \\
         \textbf{Project} & \textbf{Precision} & \textbf{Recall} & \textbf{Precision} & \textbf{Recall} \\ \hline \hline
         \textit{Stopwatch} & 0.50 & 0.8929 & 0.8837 & 1.0 \\
         \textit{DigitalHome} & 0.1257 & 0.7414 & 0.3214 & 1.0
    \end{tabular}
    \caption{Precision and recall of LLM recognized requirement flaws. Independent assessment means that both raters answered without knowledge about how the other rater answered, while bound assessment describes the situation where the study participants classified the requirements being aware of the decision and explanation of the LLM.}
    \label{tab:req_evaluation}
    \vspace{-16pt}
\end{table}

Based on our analysis, we can answer the first research question, which focuses on the accuracy of LLMs in correctly identifying quality issues in software requirements. For independent assessment, the \textit{Cohen's Kappa} metric indicates weak or even non-existent agreement between study participants and the LLM when determining whether a software requirement fulfills a quality characteristic (refer to Table \ref{tab:cohen}). However, the agreement increases for the bound assessment phase, which indicates that study participants changed their opinion toward the LLM given its explanation. We assume that the evaluation of the LLM helped correct the participants' assessment. This might help identify potential quality issues with requirements, especially when reviewers and stakeholders lack experience in RE. The LLM might partly compensate for this by suggesting additional perspectives. 

The recall values demonstrate that the LLM effectively identified the majority of flawed software requirements. This suggests it can accurately predict the presence of quality issues in software requirements (see Table \ref{tab:req_evaluation}). Nevertheless, manual verification is still required, given the low precision values, as false positives may occur. For the first research question, we conclude that LLMs can accurately identify quality flaws in many cases and show the potential to guide stakeholders toward quality issues they might have overlooked otherwise.

\subsubsection{Meaningful Explanations of LLM-Assessment (R2)}
The second research question investigates the capability of LLMs to provide reasonable explanations for their quality assessments of software requirements. To answer this question, we engaged study participants to review the decision and explanation offered by the LLM for each requirement and quality dimension. As shown in Table \ref{tab:explanation_eval}, the majority of study participants agreed that the LLM could plausibly explain its assessment of software requirement quality issues in both investigated example projects. Therefore, we can affirmatively answer the second research question. We conclude that the LLM can provide reasonable explanations for its decisions. The explanations were considered meaningful, regardless of whether the LLM classified the requirement as flawless or having a quality issue. This result is promising for the successful integration of this approach in real-life settings, as the LLM's decision is traceable for reviewers and can be accurately validated. Nonetheless, we recommend replicating the experiment with more complex requirements to observe if the LLM-generated explanations can also help generate reliable explanations for those.


\begin{table}
    \centering
    \begin{tabular}{c|c|c}
         & \textbf{Stopwatch} & \textbf{DigitalHome} \\ \hline \hline
        \textit{appropriate} & 100\% &  70\% \\
        \textit{complete}    & 100\% &  100\% \\
        \textit{conforming}  & 100\% &  100\% \\
        \textit{correct}     & 80\%  &  100\% \\
        \textit{feasible}    & 100\% &  100\% \\
        \textit{necessary}   & 100\% &  100\% \\
        \textit{singular}    & 100\% &  80\% \\
        \textit{unambiguous} & 100\% &  100\% \\
        \textit{verifiable}  & 100\% &  70\% \\
    \end{tabular}
    \caption{Percentages of LLM-generated quality-related explanations considered plausible by study participants, categorized by project and quality characteristics.}
    \label{tab:explanation_eval}
    \vspace{-16pt}
\end{table}

\subsubsection{LLM-improved Requirements (R3)}
The third research question is whether an LLM can suggest improved versions of flawed software requirements. To answer this question, we instructed study participants to evaluate the LLM-proposed improved versions of original requirements in the \textit{Stopwatch} and \textit{DigitalHome} projects. The study participants determined whether the suggested version improved the original requirement. As shown in Table \ref{tab:improvement_eval}, the suggested version was considered an improvement in most cases. As the number of requirements with potential issues with respect to feasibility and necessity was small (see Tables \ref{tab:llm-decision-dimension-requirements-stopwatch} and \ref{tab:decision-digitalhome}), there were only a few examples where an improvement could be evaluated. Therefore, more examples should be reviewed in the future to validate this aspect further.

Overall, the outcome emphasizes the LLM's capacity to assist reviewers in correcting flaws in software requirements, thereby contributing to an overall improvement in the quality of requirements. Based on the collected data, we can affirmatively answer the third research question, indicating that the LLM can suggest versions of requirements that address flaws within a given quality dimension. This shows that LLMs may offer immediate benefits in software requirements engineering.

\begin{table}
    \centering
    \begin{tabular}{c|c|c}
         & \textbf{Stopwatch} & \textbf{DigitalHome} \\ \hline \hline
        \textit{appropriate} & 100\% & 66\% \\
        \textit{complete}    & 100\% & 100\% \\
        \textit{conforming}  & 100\% & 100\% \\
        \textit{correct}     & 100\% & 100\% \\
        \textit{feasible}    & - & 50\% \\
        \textit{necessary}   & - & 50\% \\
        \textit{singular}    & 90\%  & 66\% \\
        \textit{unambiguous} & 100\% & 100\% \\
        \textit{verifiable}  & 100\% & 90\% \\
    \end{tabular}
    \caption{Percentages of LLM-generated requirement improvements considered enhancements compared to the original requirements by study participants. Values are categorized by project and quality characteristics. For \emph{feasibility} and \emph{necessity} in the \textit{Stopwatch} project, values are not available as this quality issue was not identified in the requirements and hence could not be improved.}
    \label{tab:improvement_eval}
    \vspace{-16pt}
\end{table}

\section{Discussion} \label{sec:discussion}
\subsection{Study Results}
The results of this empirical study provide valuable insights into the potential and capabilities of leveraging an LLM to support the development of high-quality software requirements. As demonstrated by the results, the LLM successfully identified a majority of the incorporated quality issues, showing its utility in the requirements review process. This ability can significantly reduce review time by directing reviewers to requirements that may contain quality flaws. However, it is important to note that the intervention of the reviewers remains necessary, as the LLM did not identify all quality issues in the examined samples. The level of collaboration between humans and LLMs is a topic of recent research. Faggioli et al. \cite{faggioli2023perspectives} outline various integration stages from full human judgment to complete automation. We assume that LLMs are currently most beneficial in assisting humans by suggesting and explaining potential flaws for the quality assurance of software requirements. Those need to be subsequently verified and corrected with the help of the LLM. The current capabilities of LLMs are not sufficient to fully automate this task.

The LLM consistently offered meaningful explanations, regardless of whether participants agreed with the LLM's quality assessment. This suggests that the LLM is transparent in articulating its reasoning, potentially helping reviewers understand why a requirement was identified as having a quality issue. Additionally, these explanations may encourage reviewers to revise their assessments by considering an alternative perspective. We believe that explaining decisions is crucial for establishing LLM-assisted review processes for software requirements. Such explanations could facilitate the validation of assessments and increase trust in the system.

Moreover, the recommendation of improved requirements appears beneficial and reliable based on the evaluation of our examples. The study participants agreed that most of the LLM-suggested requirements improved their quality, regardless of whether the study participants agreed with the LLM's quality assessment. This additional perspective might help reviewers by providing an alternative viewpoint and potentially refining the formulation of requirements where quality issues are not particularly severe. In this study, the LLM was explicitly instructed to correct the identified flaw concerning the assessed quality characteristic, ignoring the possibility that a requirement might require improvement across multiple dimensions. Addressing the combination of different dimensions in a single prompt instruction to achieve an overall enhanced requirement remains an important consideration for future research.

Overall, we believe that LLMs have the potential to enhance the quality of software requirements, particularly by reducing the manual review effort. It is crucial to interpret the findings of our study as a proof-of-concept, emphasizing the need to develop and evaluate ideas regarding this application further.

\subsection{Threats to Validity}
While the study presented in this paper provides valuable insights into the potential of assuring and improving the quality of software requirements with LLMs, we acknowledge certain limitations. \emph{Firstly}, the evaluation was conducted using a hypothetical example project and only one real-life scenario, impacting the generalizability of the presented results. To obtain better evidence of the general applicability, we plan to examine the presented approach within upcoming real-life projects, using the LLM as a supportive tool during the review phase of requirements. With this, we hope to learn more about its usefulness in practical settings.

\emph{Secondly}, the investigated projects were relatively simple, meaning that the reported results might differ for larger projects with more complex requirements. Additionally, we did not address the token limit aspect in the prompts for more extensive project descriptions. In such cases, a \textit{Retrieval Augmented Generation (RAG)} \cite{lewis2020retrieval,gao2024retrievalaugmented} approach could be employed to overcome this challenge by loading relevant project data into the prompt context as needed to reduce the context length.

\emph{Thirdly}, our study did not explore and compare different options to instruct LLMs. The focus was on one specific model (Llama 2) and a single prompt template, making it impossible to estimate the general quality of LLMs in evaluating software requirements. The usage of different LLMs or prompts might lead to other results. Consequently, we consider future studies addressing this aspect by comparing various prompting strategies and models. Given the extensive research on LLMs and the ongoing release of improved model versions, it is important to continuously evaluate new approaches for interaction and application in experimental studies.

\subsection{Future Work}
In addition to assessing the quality of individual requirements, we see the potential for applying our approach to evaluate the quality characteristics of requirement sets. Similar to the quality dimensions for individual statements, the \textit{ISO 29148} \cite{iso29148} specifies characteristics for sets or requirements, encapsulating the quality aspects of a consistent solution that meets stakeholder expectations while adhering to project constraints. We believe that extending our approach to cover these aspects could yield additional benefits. 

Additionally, we consider the potential of using an LLM to identify hidden dependencies between requirements as an open issue. Existing approaches for detecting these dependencies \cite{atas2018automated,samer2019new} use natural language techniques to extract features as a basis for classification of dependencies between requirements. In our future research, we aim to develop a framework for instructing an LLM to handle this task. We believe that the extensive general knowledge of LLMs, combined with their ability to understand specific project contexts, could significantly enhance the accuracy of identifying these dependencies.

Furthermore, we intend to analyze additional projects to provide a more general understanding of the capabilities of LLMs in improving software requirements. For this, we consider the application of \textit{Retrieval Augmented Generation} \cite{lewis2020retrieval,gao2024retrievalaugmented} for complex and huge projects as a promising approach for future research. In this context, exploring strategies to identify and access the relevant information is essential to let the LLM understand the project scope.

\section{Conclusions} \label{sec:conclusion}
In this paper, we analyzed the possibilities of using an LLM to evaluate the quality of software requirements in accordance with the quality characteristics defined in the \textit{ISO 29148} standard. We discuss an approach for instructing the LLM to assess requirements, explain its decision-making process, and suggest enhanced versions when quality issues are detected. To evaluate the LLM's capabilities in fulfilling these tasks, we conducted an empirical study involving participants with backgrounds in software engineering. The findings indicate that the LLM accurately identifies the majority of requirements with quality flaws, demonstrating its potential to assist in the requirement engineering process. Moreover, the study reveals that the LLM provides reliable explanations for its decisions, which can improve trust in the system. The evaluation of LLM-suggested improvements to requirements with quality issues indicates their effectiveness in resolving quality concerns. To summarize, this paper highlights the utility of LLMs in ensuring the quality of software requirements, suggesting promising potentials for stakeholder support throughout the software requirements engineering process, and motivating future related research.


\bibliographystyle{IEEEtran}
\bibliography{bibliography}

\end{document}